\documentclass[twocolumn,showpacs,preprintnumbers,amsmath,amssymb]{revtex4}

\usepackage{graphicx}
\usepackage{dcolumn}
\usepackage{bm}

\begin{document}


\title{Fluctuating order parameter in doped cuprate superconductors}

\author{V.M. Loktev}
\altaffiliation{Electronic address: vloktev@bitp.kiev.ua}
\affiliation{Bogolyubov Institute for Theoretical 
Physics, Metrologichna str. 14-b, Kiev, 03143 Ukraine}

\author{Yu.G. Pogorelov}
\altaffiliation{Electronic address: ypogorel@fc.up.pt}
\affiliation{CFP and Departamento de Fisica, Faculdade de
Ciencias, Universidade do Porto, Rua do Campo Alegre, 687,
4169-007 Porto, Portugal}

\author{and V.M. Turkowski}
\altaffiliation{Electronic address: vturk@cfif.ist.utl.pt}
\affiliation{CFIF, Instituto Superior Tecnico,
Av.Rovisco Pais, 1049-001 Lisbon, Portugal}%

\date{\today}

\begin{abstract}
We discuss the static fluctuations of the d-wave superconducting order 
parameter $\Delta$ in CuO$_2$ planes, due to quasiparticle scattering 
by charged dopants. The analysis of two-particle anomalous Green functions 
at $T = 0$ permits to estimate the mean-square fluctuation $\delta^2 = 
\langle\Delta^2\rangle-\langle\Delta\rangle^2$, averaged in random dopant 
configurations, to the lowest order in doping level $c$. Since $\Delta$ 
is found to saturate with growing doping level while $\delta$ remains to 
grow, this can explain the collapse of $T_c$ at overdoping. Also we 
consider the spatial correlations $\langle\Delta(0)\Delta({\bf R})\rangle$ 
for order parameter in different points of the plane.
\end{abstract}

\pacs{74.20.-z, 74.62.Dh, 74.72.-h}


\maketitle

\section{\label{int} Introduction}
Recently, a new piece emerged that physicists should insert into the 
puzzle of high-$T_c$ superconductivity (HTSC) in metal-oxide materials. 
The advances in STM techniques permitted to obtain nanoscopic resolution 
of their electronic structure, revealing a non-uniform pattern of local 
density of states and of superconducting (SC) d-wave gap.\cite{1},\cite{2} 
For instance, in underdoped Bi$_2$Sr$_2$CaCuO$_{8+x}$ the islands of 
$\sim 3$ nm size of almost saturated SC order (sharp coherence peaks in 
the density of states) are surrounded by undersaturated areas (low peaks); 
the pattern also remains non-uniform in the optimally doped material but 
with undersaturated islands over the saturated background.\cite{2} This 
evidence is corroborated by the results of precision STM experiments on 
HTSC in the mixed state under applied magnetic field $H \ll H_{c2}$, where 
no common lattice order for Abrikosov vortices was found.\cite{3},\cite{4} 
This experimental challenge invoked a massive theoretical 
response,\cite{5}-\cite{10} in particular, the scenario of two distinct 
(and competing) SC phases was set forth,\cite{1},\cite{5}-\cite{7} 
supposing segregation of an undersaturated $\beta$-phase from the saturated 
$\alpha$-phase and associating it with nucleation of the pseudogap state or 
the so-called d-density wave state.\cite{11} However, this way of reasoning 
seems to lead to even harder problems with microscopic mechanisms for such 
strong and irregular non-uniformity and it is in odds with the concept of 
stripe phases (another piece of the HTSC puzzle \cite{12},\cite{13})

Perhaps the most realistic approach to the problem is that based on the 
evident attraction of charge carriers in HTSC to each other and to the 
ionized dopants.\cite{8}-\cite{10},\cite{14},\cite{15} These two factors 
are competitive: the SC coupling $V$ leads to formation of uniform states 
of correlated pairs of charge carriers, while the impurity potential $V_i$ 
leads to formation of localized states of single carriers and to local 
reduction of pairing amplitude.\cite{15},\cite{16} Taking into account that 
the number of ionized dopants equals the number of carriers, the "impurity" 
effect can be well competitive with the pairing for all the levels of doping. 
Below we use the method of Green functions to estimate the impurity effect 
on SC properties of doped metal-oxides, especially the amplitude and spatial 
correlations of fluctuations of the d-wave order parameter.

\section{\label{sp} Green functions and single-particle properties}
We start from the model Hamiltonian:

\begin{eqnarray}
H = \sum_{\bf k} \psi^\dagger_{\bf k}(\xi_{\bf k}\hat\tau_3+
\Delta_{\bf k}\hat\tau_1)\psi_{\bf k} \qquad \notag \\
+\frac{V_i}{N}\sum_{{\bf k},{\bf k}^\prime,{\bf p}} 
{\rm e}^{i({\bf k}-{\bf k}^\prime){\bf p}} \psi^\dagger_{\bf k}
\hat\tau_3\psi_{{\bf k}^\prime}\,, \label{eq1}
\end{eqnarray}

\noindent where the Nambu spinors $\psi^\dagger_{\bf k}=(a_{{\bf k},
\uparrow},a^\dagger_{-{\bf k},\downarrow})$ include fermion operators 
of normal quaisparticles with the 2D dispersion law $\xi_{\bf k} = 2t
(2 - \cos ak_x - \cos ak_y)-\mu$, the bandwidth $W = 8t$, and the 
chemical potential $\mu$; and $\hat\tau_i$ are the Pauli matrices. 
The gap function $\Delta_{\bf k} = \Delta\gamma_{\bf k}\theta
(\varepsilon_D^2-\xi_{\bf k}^2)$ includes the BCS-shell restriction 
by the "Debye energy" $\varepsilon_D$, and the d-wave factor $\gamma_
{\bf k} = \cos ak_x - \cos ak_y$. In the long-wave limit: $\gamma_{\bf k}
\approx 1/2(ak)^2\cos 2\phi_{\bf k}$ ($\phi_{\bf k}$ is the azimuthal 
angle in $k$-plane), hence the observable coherence peak should lie at 
$\Delta_{obs}\approx \pi \mu \rho_0 \Delta$, where $\rho_0 = 4/(\pi W)$ is 
the constant density of states of normal metal in this limit. The 
impurity perturbation $V_i$ produces scattering of quasiparticles at 
lattice points ${\bf p}$ randomly distributed with the doping 
concentration $c$. The true electronic spectrum follows from the Fourier 
transform of two-time fermion Green functions (GF's)

\begin{equation}
\langle\langle a|b\rangle\rangle_\varepsilon = i\int^\infty_0 {\rm e}
^{i(\varepsilon+i0)t}\langle\{a(t),b(0)\}\rangle dt\,, \label{eq2}
\end{equation}

\noindent (where $\langle\ldots\rangle$ is the quantum statistical average 
with the Hamiltonian \ref{eq1} and $\{\ldots,\ldots\}$ is the anticommutator 
of Heisenberg operators), through the spectral formula:

\begin{equation}
\langle a b \rangle = \frac{1}{\pi}\int^\infty_{-\infty} \frac{d
\varepsilon{\rm Im}\langle\langle b|a\rangle\rangle_\varepsilon}
{{\rm e}^{(\varepsilon-\mu)/T}+1}\,. \label{eq3}
\end{equation}

\noindent Thus, the Nambu matrix $\hat G_{\bf k} = \langle\langle 
\psi_{\bf k}|\psi^\dagger_{\bf k}\rangle\rangle$ of momentum-diagonal 
single-particle GF's and its local value $\hat G = N^{-1}\sum_{\bf k} 
\hat G_{\bf k}$ enter the two coupled equations\cite{17}

\begin{eqnarray}
c=1+\frac{1}{\pi}\int^\mu_{-\infty} d\varepsilon{\rm Im}{\rm Tr}
\hat G \hat \tau_3 \, ,\qquad \qquad \notag\\ 
\Delta = \frac{V}{2\pi N}\sum_{\bf k} \gamma_{\bf k} 
\int^\mu_{-\infty}d\varepsilon{\rm Im}{\rm Tr}
\hat G_{\bf k}\hat\tau_1\,, \label{eq4}
\end{eqnarray}

\noindent which define the chemical potential $\mu$ and the gap parameter 
$\Delta$ in function of the doping level $c$ and of the parameters $V$, 
$V_i$. The standard analysis\cite{14} gives $\hat G_{\bf k} =
(\varepsilon-\mu-\xi_{\bf k}\hat\tau_3 - \Delta_{\bf k}\hat\tau_1 -
\hat\Sigma_{\bf k})^{-1}$, where the self-energy matrix can be approximated 
in the self-consistent t-matrix form $\hat\Sigma_{\bf k} \approx c\hat 
V (1-\hat G \hat V)^{-1}$ with $\hat V = V_i\hat\tau_3$, then the 
integrals in Eq. \ref{eq4} yield in the approximate analytic formulae:

\begin{eqnarray}
\mu \approx \frac{c-c_{met}}{\rho_0}\,,\qquad \qquad \qquad \qquad \notag \\
\Delta_{obs} \approx \frac {2 \varepsilon_D}{\pi} \exp 
\left\{-\frac{1}{(\pi\mu\rho_0)^2} \left[ \frac{2}{V\rho_0} 
- \left(\frac{\pi\varepsilon_D\rho_0}{2}\right)^2
 \right] \right\} \,. \label{eq5}
\end{eqnarray}

\noindent In fact, the first formula in Eq. \ref{eq5} only applies for 
doping levels $c$ well above the metallization threshold $c_{met} \approx 
\varepsilon_0 \rho_0 /2$, where $\varepsilon_0 = \{\rho_0[\exp(1/V_i\rho_0)
-1]\}^{-1}$ is the binding energy of impurity localized state. For $c < 
c_{met}$, the system is insulating (so non-superconducting) but the chemical 
potential remains positive, vanishing as $\sim c/[\rho_0\ln(c_{met}/c)]$ at 
$c\to 0$. The second formula strictly applies at $c > \varepsilon_D
\rho_0$, while at $c$ below this value the pre-exponential factor should 
be changed for $2\sqrt{\varepsilon_D\mu}/\pi$. A more detailed discussion 
of the averaged single-particle spectrum will be given elsewhere.

\section{Two-particle properties and fluctuations of SC order}
\label{tp}
The non-uniform SC order can be described, using the operator of local 
d-wave order  on ${\bf n}$th lattice site: $\Delta_{\bf n} = (V/N)
\sum_{{\bf k},{\bf k}^\prime} \gamma_{\bf k} {\rm e}^{i({\bf k}-{\bf k}
\prime){\bf n}} a_{{\bf k},\uparrow} a_{-{\bf k}^\prime,\downarrow}$, with 
the average value $N^{-1}\sum_{\bf n}\langle\Delta_{\bf n}\rangle = 
\Delta$. Its squared dispersion $\delta^2 = N^{-1}\sum_{\bf n}
(\langle\Delta^2_{\bf n}\rangle-\langle\Delta_{\bf n}\rangle^2)$ is 
written from Eq. \ref{eq3} as:

\begin{eqnarray}
\delta^2 =\frac{V^2}{N^2}\sum_{{\bf k},{\bf k}^\prime,{\bf q}}
\gamma_{\bf k} \gamma_{{\bf k}^\prime} \times \qquad \qquad \qquad \notag \\
\times \left( \frac{1}{\pi} 
\int^\mu_{-\infty} d\varepsilon {\rm Im} \langle\langle 
a_{{\bf k},\downarrow}a_{-{\bf k}^\prime,\uparrow}|a_{-{\bf k}^\prime 
+{\bf q},\downarrow}a_{{\bf k}-{\bf q},\uparrow} \rangle\rangle \right.
\nonumber \\
- \frac{1}{\pi^2}\int^\mu_{-\infty} d\varepsilon{\rm Im}\langle\langle 
a_{{\bf k},\downarrow}|a_{-{\bf k}^\prime,\uparrow}\rangle\rangle \times
\qquad \qquad \notag \\
\times \left. \int^\mu_{-\infty} d\varepsilon{\rm Im}\langle\langle 
a_{-{\bf k}^\prime +{\bf q},\downarrow}|a_{{\bf k}-{\bf q},\uparrow} 
\rangle\rangle \right)\,. \label{eq6}
\end{eqnarray}

\noindent It is zero in the doped but uniform crystal (at $c > 0$ but 
$V_i = 0$), whereas the non-zero contribution to Eq. \ref{eq6} in the 
lowest order in $V_i$ comes from two successive scatterings: ${\bf k}\to 
{\bf k} + {\bf q}$, ${\bf k}^\prime \to {\bf k}^\prime - {\bf q}$, involved 
into the equations of motion for the momentum non-diagonal two-particle 
(anomalous) GF's. In this approximation, we have

\begin{equation}
\delta^2 \approx \frac{c V^2_i V^2 \Delta^2}{2} \sum_{{\bf k},{\bf k}^\prime}
\gamma^2_{\bf k} \gamma^2_{{\bf k}^\prime}\frac{\theta(E_{\bf k} -
E{{\bf k}^\prime})}{E_{\bf k}E^3_{{\bf k}^\prime}}\approx
\frac{cV^2_i V\rho_0}{4}\,. \label{eq7}
\end{equation}

\noindent Accordingly to the written in Sec. \ref{sp}, the dispersion of 
the observable gap is $\delta_{obs} = \pi\mu\rho_0\delta \approx (\pi/2)
\sqrt{V\rho_0}(c-c_{met})^{3/2}V_i$, and comparing it with Eq. \ref{eq5}, 
we conclude that this dispersion turns comparable with the average gap and 
can bring the SC order to collapse at $c \sim (\varepsilon_D/V_i)^{2/3}
(V\rho_0)^{-1/3}\exp(\pi^4 V\varepsilon_D\rho^2_0/12)$.

Finally, we can estimate the spatial correlation between the above 
mentioned fluctuations, alike Ref. \cite{10}, from the analysis of the 
random function $f({\bf n}) = \sum_{\bf m} c_{\bf m}\exp[-({\bf m}-
{\bf n})^2/\xi^2_0]$, where $c_{\bf m} = 1$ for dopant perturbed sites, 
${\bf m} = {\bf p}$, otherwise $c_{\bf m} = 0$, and $\xi_0$ is the BCS 
coherence length. Then the spatial decay of the correlator of fluctuations 
$\langle f({\bf n})f({\bf n}\prime)\rangle - \langle f({\bf n})\rangle \langle 
f({\bf n}\prime)\rangle \approx c(1-c)\exp[-({\bf m}- {\bf n})^2/2
\xi^2_0]$ defines their correlation length $\xi_c \approx \sqrt{2}
\xi_0$. This implies an estimate for the average pinning force due to 
such non-uniform SC structure $F_p \sim \rho_0\delta_{obs}\Delta_{obs}
(\xi_0/a)^2/\xi_c$. Comparison of this and other predictions with the 
available experimental data will be the subject of forthcoming studies.

\section*{Acknowledgments}

The authors acknowledge the financial support from Swiss Science 
Foundation under the Project SCOPES 7UKPJ062150.00/1 (V.M.L.) and from 
Portuguese Funda\c{c}\~ao de Ci\^encia e Tecnologia under the Projects 
POCTI/1999/CTM/36489 (Yu.G.P.) anf FJ08 (V.M.T.).

\end{document}